**Bio-inspired adaptive sensing through electropolymerization of organic electrochemical transistors**

*Mahdi Ghazal, Michel Daher Mansour, Corentin Scholaert, Thomas Dargent, Yannick Coffinier, Sébastien Pecqueur*, and Fabien Alibart**

M. Ghazal, Dr. M. Daher Mansour, C. Scholaert, Dr. T. Dargent, Dr. Y. Coffinier, Dr. S. Pecqueur, Dr. F. Alibart
Institut d'Électronique, Microélectronique et Nanotechnologies (IEMN) – CNRS UMR 8520 – Université de Lille, boulevard Poincarré, 59652 Villeneuve d'Ascq, France.
E-mail: sebastien.pecqueur@univ-lille.fr; fabien.alibart@univ-lille.fr;

Dr. F. Alibart
Laboratoire Nanotechnologies Nanosystèmes (LN2) – CNRS UMI-3463 – 3IT, Sherbrooke J1K 0A5, Canada

**Abstract:** Organic Electrochemical Transistors are considered today as a key technology to interact with biological medium through their intrinsic ionic-electronic coupling. In this paper, we show how this coupling can be finely tuned (in operando) post-microfabrication via electropolymerization technique. This strategy exploits the concept of adaptive sensing where both transconductance and impedance are tunable and can be modified on-demand to match different sensing requirements. Material investigation through Raman spectroscopy, atomic force microscopy and scanning electron microscopy reveals that electropolymerization can lead to a fine control of PEDOT microdomains organization, which directly affect the iono-electronic properties of OECTs. We further highlight how volumetric capacitance and effective mobility of PEDOT:PSS influence distinctively the transconductance and impedance of OECTs. This approach shows to improve the transconductance by 150% while reducing their variability by 60% in comparison with standard spin-coated OECTs. Finally, we show how to the technique can influence voltage spike rate hardware classification with direct interest in bio-signals sorting applications.

1. Introduction

Organic Electro-Chemical Transistors (OECTs) have emerged as a key technology in both bio-sensing and neuromorphic engineering.[1–3] As bio-sensors, OECTs offer very attractive solutions for converting ionic signals into electronic ones thanks to the unique property of organic mix ionic electronic conductors (OMIECs).[4] Ionic concentration from an analyte or ionic currents from electroactive cells can be efficiently sensed/probed and amplified, thus making OECTs attractive sensors.[5] In the perspective of neuromorphic engineering, the same devices are capitalizing on the possibility to engineer devices where ion-electron coupling can be used to implement various synaptic plasticities, from short-term to long-term memory effects.[3,6–9] These two aspects have been so far mostly developed independently from each other. In contrast, synapses in biology are combining sensing capabilities with plastic properties to provide some essential aspects of bio-computing. Through their adaptation properties, synapses are enhancing/depressing relevant/irrelevant signals from neurons. They also provide a rich set of non-linear operations to process the spike signals from neural cells.[10] As sensors, synapses are converting chemical signals from sensed neurotransmitters into transduced post-synaptic electric signals as ionic concentration modulation. Such ambivalence existing in biology is the natural example of a non-Von Neuman computing architectures that embeds highly complex biochemical sensing at all nodes in its network, and demonstrates reciprocally the power of local adaptation of a sensing array that programs according to its environment.

In this paper, we show how OECTs can combine these two important features for bio-signal sensing and processing. The corner stone of OECTs behavior is the transconductance, which couple ionic signals to electronic ones.[11] Transconductance can be well described by the coupling between (i) volumetric ionic capacitance allowing for very large effective surface of interaction between the analyte and the polymer and (ii) efficient electronic transport along the π-conjugated organic chains. Several works have demonstrated routes for optimizing tranconductance through either volumetric capacitance or electronic mobility tuning.[12,13] Here, we show how electropolymerization can be used to adapt post-fabrication these two intrinsic parameters of OECTs and how this technique results in bio-inspired adaptive sensing. We combine standard spin-coating deposition technique with post-fabrication potentiostatic electropolymerization (EP) to show how OMIECs properties can be adjusted finely. Organization of the PEDOT:PSS domains are revealed by scanning electron microscopy (SEM) and atomic force microscopy (AFM) for different EP conditions. Raman spectroscopy is used to highlight π-conjugated domains organization and redox state of the PEDOT under different deposition conditions. We combine DC electrical measurements with electrochemical impedance spectroscopy to extract volumetric capacitance and electronic mobility. These material aspects are then employed to show how OECTs' transconductance from a multi-sensor array can be adapted on-demand to an optimal transconductance value, which mimic the long-term potentiation process in biological synapses. We further demonstrate how tuning OECTs intrinsic properties can result in tunable frequency filters, which provides a unique opportunity for processing spike-based signals equivalently to short-term plasticity effects as a memory mechanism enabled in biological synapses.

2. Results and discussion

OECTs were realized following a conventional route involving Au source (S) and drain (D) electrodes patterning through optical lithography and lift-off. A two layers parylene C was subsequently used to define a 30 µm diameter opening on top of the S and D contact. The resulting W/L for OECTs was 30/12. The first layer was used for contact line insulation while the second one acts as a sacrificial layer. PEDOT:PSS (Clevios$^{TM}$ PH1000 ) was first spin-coated before sacrificial layer removal resulting in a 200 nm thick spin-coated PEDOT:PSS in between S and D contacts (figure 1). This standard process for OECTs fabrication results in well-established transconductance performances $G_m \propto \mu.C^*$. Hole mobility $\mu$ is the effective mobility of electronic carriers along the π-conjugated system and results from both delocalized states (i) intromacromolecular π-σ-π electron delocalization along the thiophene backbone or (ii) hoping from π-π -stacked intermolecular thiophene rings (iii) intermolecular hoping between two PEDOT macro-molecules that are adjacent from a same PSS macromolecular chain.[14] Intermolecular hoping is the most limiting mechanism that largely defines the effective mobility of electronic carriers through the material. Volumetric capacitance C* is describing the permeability of PEDOT:PSS to ions allowing positively charged ions to interact with PSS⁻ fixed charges in the bulk of the material.[15,16] Direct correlation between Gm and both μ and C* justifies why transconductance is the main figure of merit describing both ionic and electronic properties of OMIECs.[17]

It turns out that improving transconductance has been realized following either volumetric capacitance or electronic mobility tuning. For spin-coated PEDOT:PSS, improving volumetric capacitance has been realized by material engineering routes consisting in addition of hydrophilic chains and molecules[18,19] to the spin-coated solution that results in larger volumetric capacitance values of up to 6-57 F/m3. A more straightforward solution is to increase the total thickness of PEDOT:PSS in order to increase the total capacitance of the OECT.[20] This technique has resulted in record transconductance of ~4000 μS, surpassing 2D materials mobility such as graphene.[1] Nevertheless, as a soft fabrication process, spin-coating by itself does not allow adjusting the volumetric capacitance. Successive spin-coating steps are required and lead to a poor control over the total material thickness and variability. This holds also for the control of the effective mobility (i.e. organization of the π-conjugated system in PEDOT:PSS) that is mostly defined by chemical interaction of the polymer in its solution and its kinetics for drying upon a specific process. Notably, PEDOT molecules have been reported to organize along the PSS chains resulting in fiber-like structure.[14] Addition of heterogeneous elements in the solution have demonstrated improve mobility of spin-coated PEDOT:PSS.[21] Recently, effective mobility improvement was obtained by stress-engineering in PEDOT-PSS fibers resulting in improve organization of the π-conjugated system along the fiber with effective mobility as high as 12.9 $cm^2.V^{-1}.s^{-1}$.[22] It turns out that both capacitance and effective mobility improvement for high tranconductance are only accessible via various material synthesis and engineering techniques that cannot be necessarily combined. This lack of engineering flexibility represents a strong limitation for future development of OECTs applications.

2.1. Microstructural analysis of electropolymerized PEDOT:PSS

In our approach, we propose to combine spin-coating with electropolymerization in order to gain an additional level of freedom in adjusting ionic and electronic properties of OECTs. Electropolymerization (EP) technic has been largely investigated for PEDOT materials deposition.[23,24] Here, EP of PEDOT:PSS was realized in a potentiostatic configuration as shown in Figure 1d with S and D as working electrode ($V_{IN}$) and grounded Pt wire as reference electrode ($V_{OUT}$) dipped into the electrolyte that contained the monomer (0.01 M EDOT + 0.1 M PSSNa). EP was realized on top of a standard spin-coated layer of a PEDOT:PSS (see experimental methods section). A fiber-like structure was obtained for spin-coated PEDOT and granular structure was obtained for electropolymerized materials as shown in the SEM and AFM images of Figure 2. Statistical analysis of these AFM images was realized to reveal the grain size, surface roughness and thicknesses of the electropolymerized thin films. For 0.6 V EP potential, AFM images showed granular growth with a grain size of ~5 nm whereas grain size of ~9 nm were formed for 0.7 V potential. Surface average roughness increased from ~3 nm, ~ 4 nm and ~8 nm for the spin-coated, 0.6 V and 0.7 V electropolymerized PEDOT films, respectively. A clear relation between grain size and roughness can be drawn from these measurements with similar ratios of ~ 1:2 for both 0.6 V and 0.7 V EP potentials. Moreover, height profiles of AFM images before and after EP revealed electropolymerized thin films of (100 ± 4) nm and (200 ± 8) nm for 8 s of EP at 0.6 V and 0.7 V, respectively. The thickness of the two thin films presented a ratio of 1:2, which might explain the increase of the roughness and the creation of less organized surface for higher deposition rate. More interestingly, EP gave us access to various thin films microstructures that should relate to different electrical properties of the materials.

Figure 2j shows the Raman spectra of PEDOT:PSS grown by spin-coating (black line) and by EP at 0.6 V (green line) and at 0.7 V (blue line) potentials. These spectra are typical of the ones shown in the literature for PEDOT:PSS.[25–27] The bands assignment are presented in Table S1. From these measurements, we identified different orientations of polymers deposited on the surface and doping levels among the three different deposition conditions. Concerning the doping level, the peak for the $C_\alpha=C_\beta$ symmetric vibrations shifted from 1439 cm$^{-1}$ for the spin-coated PEDOT to 1444 cm$^{-1}$ for both electropolymerized PEDOT. This shifts indicates a higher level of oxidation for the polymer[25,27–32] in correlation with the peaks at 1266 cm$^{-1}$ that corresponded to the oxidized state of PEDOT.[25,27–32] In addition, these blue-shifts (i.e. to higher wavenumbers) could be associated to oxidized states with quinoid form of polymer chains, which may have a key role on the conductivity of the polymers.[30] Moreover, electropolymerized PEDOT at 0.6 V presented the narrowest peak at 1444 cm$^{-1}$ which implies a longer degree of conjugated length and thus higher crystallinity and conductivity as shown by Zhao et al. [28],[33] On the other hand, the decrease in peaks' intensity in the 400-1000 cm$^{-1}$ region (Oxyethylene ring deformation) whereas an increase is observed at 1450-1650 cm$^{-1}$ ($C_\alpha=C_\beta$ vibrations region for both electropolymerized PEDOT relatively to spin-coated PEDOT indicated a perpendicular orientation of the thiophene rings with the surface and thus an edge on growth of PEDOT:PSS on the surface.[29,34–36] To complete the discussion on the Raman measurements, electropolymerizied PEDOT at 0.6 V and 0.7 V showed different intensity peaks in the 1300-1600 cm$^{-1}$ region for the $C_\beta-C_\beta$ stretching and the $C_\alpha=C_\beta$ asymmetric vibrations. These differences in relative intensities also indicated a different structure orientation of the polymer chains with the surface, and thus different electrical properties. As illustrated in the schematic of Figure 2k, the relatively perpendicular thiophene rings organization leads to an in-plane π-π stacking which in turn increases the mobility in the thin films. In agreement, narrower peaks for PEDOT at 0.6 V indicates better crystallinity and thus better conductivity. Wu and coworkers[36] showed that a π-π stacking order in a lamella interchain stacking structure for an edge-on PEDOT orientation increased the conductivity of the thin film polymer. Therefore, we can expect here a better crystallinity structure for PEDOT grown at 0.6 V with the latter ordering, which led to a better conductivity of the thin film polymer.

From a structural and organizational standpoint, the Raman and AFM results correlated. The results were in agreement with a smoother and more organized PEDOT:PSS chains for the 0.6 V EP. These different growth characteristics (thickness, organization) revealed by AFM and Raman might give us a hint on the electrical properties of the PEDOT:PSS thin films. The growth at 0.6 V potential is expected to present a better conductivity while the growth at 0.7 V is expected to present higher capacitance. These findings suggest that electrical properties can be tuned by applying different conditions of electropolymerization. Note that the choice of Pt as a reference electrodes is defining the operating EP potential. Other reference electrodes should result in different EP potentials but the methodology of our approach should remain valid.

## 2.2. Transconductance and impedance evaluation

Combination of spin-coating with EP is offering an interesting option for accessing to different material microstructures. Different microstructures are affecting the transconductance through either a change in capacitance or in electronic mobility. To access to the ionic and electronic properties of the film, we conducted both DC electrical characterization of the OECTs and electrochemical impedance spectroscopy for various EP conditions. EP was realized sequentially on the same OECTs by step of 2 seconds. Transfer characteristics and impedance spectrum were recorded between each steps. Figure 3 presents the evolution of transfer characteristic and impedance for potentiostatic EP on top of spin-coated PEDOT:PSS at $V_p$ = 0.6 V. Transfer characteristics provide a direct access to transconductance through $G_m$= $dI_{ds}/dV_g$, with $I_{ds}$ the channel current and $V_g$ the gate voltage. Since thicknesses after EP were not measured, we cannot extract conductivity and mobility from DC measurements. Impedance spectroscopy was modeled with an equivalent electrical circuit (figure 3c) giving access to the total capacitance of the OECT. A clear increase of transconductance was observed as well as a clear increase in the capacitance (i.e. shift of the -1 slope part of the impedance modulus spectrum below 1 kHz). We conducted the same experiment for various potentiostatic conditions. Figure 3d presents the relative change of maximum transconductance as a function of the relative change of capacitance. Following the work of [20], the dashed line in figure 3d presents the linear relation expected in the case of a simple increase of material thickness. In this scenario, an increase of $G_m$ is solely related to an increase of volume leading to higher capacitance. Interestingly, we observed that 0.6 V EP condition led to an evolution above this unity line while larger potential led to an evolution below it. These trends indicate that low EP potential allows to grow material with better mobility than spin-coated one and high potential favor lower mobility materials. In other words, EP provides a versatile technique that can be used to tune both electronic mobility and ionic capacitance. Notably, a 150% increase in the peak transconductance with potentiostatic EP of PEDOT:PSS OECTs is demonstrated. We note that transconductance is dependent on capacitance, mobility and threshold voltage ($V_{th}$). Our analysis is disregarding the effect of $V_{th}$. Figure S4 presents the dependency of $G_m$ with both capacitance and $V_{th}$. A weak dependency of $G_m$ with $V_{th}$ was observed for a change in $G_m$ with constant capacitance, thus pointing in the direction of the contribution of mobility.

**2.3. Capacitance and effective mobility tuning for adaptive OECTs sensors.**

Versatility of electropolymerization is an attractive solution to engineer post-fabrication ionic and electronic properties in OMIECs. Having independent control over OMIECs capacitance and mobility can be used to adapt OECTs responses depending on the targeted application. In biological neural networks, synapses are experiencing long-term modification of their strength through learning to process efficiently bio-signals. This property is extensively used in artificial neural network through supervised training of networks to implement fundamental tasks such as signals classification or features extractions. Equivalently, tunable OECTs sensors with adaptive transconductance could be used to extract important features from the biological medium by weighting the transconductance of each sensor individually. Figure 4a presents the evolution of peak transconductance when successive electropolymerization were realized to reach a 1 mS tranconductance value. In our experiment, the transconductance was recorded after every 2 seconds of EP. When tranconductance of more than 1 mS was reached or EP time was larger than 10 s, the tranconductance tuning was stopped. In addition to a mean transconductance value of 0.98 mS (i.e. 2% deviation from the targeted value) a large decrease of standard deviation in transconductance from 0.24 to 0.086 mS was obtained for a batch of 30 devices belongings to the same array. Figure 4b shows the evolution of the capacitance for the same experiment. The low initial variability in capacitance associated to the large initial variability in transconductance of spin-coated PEDOT:PSS was pointing that variability in mobility rather than variability in capacitance was the initial source of variability in transconductance. The increase in capacitance variability after EP was suggesting that variability in transconductance was reduced by adjusting capacitance rather than mobility. It is to note that tranconductance tuning could be optimized further by using automated setup that could allow fine-tuning over time by combining EP and tranconductance measurements in the same environment. In our case, the setup was limited by the manual operation of the setup. Also, this demonstration implemented only long-term potentiation effects (i.e. increase of the transconductance). As suggested by the decay of transconductance for long EP time in figure 3d, long-term depressions (LTD) should be also available. Origin of LTD could be explained by either (i) a degradation of mobility of the previously deposited material by decreasing thiophene rings organization when more material was added or (ii) ions diffusivity limitation when too thick materials are deposited. Further investigations are required to conclude on this aspect.

A second bio-inspired feature available with adaptive OECTs is to implement Short Term Plasticity (STP). Here we capitalized on the bottom-up tuning of OECTs capacitance to show how an additional degree of freedom can be obtained with EP. Figure 5 compares the response to a train of 100 µs pulses at 1 kHz and 8 kHz before and after EP. Figure 5b shows the response of the OECT with partial charging during the 200 mV pulse and discharge at 0 V. STP effect results from the balance in between the charging and discharging of ions during these two phases. After EP at 0.7 V, the capacitance of the OECT was increased from 10 nF to 19 nF, thus increasing the time constant of the equivalent Resistor-Capacitor circuit. Before EP, OECT's response presented higher modulation than after EP at low frequency. Note that in this low frequency regime, charge and discharge were completely balanced. At high frequency, both device reached an equivalent saturation regime corresponding to the maximal accumulation of positive ions. In this later case, charging was not balanced by discharging due to short inter-pulses interval and current reached the saturation level. Comparison in between the two responses demonstrates the possibility to tune the high-pass filtering property of the OECT with EP. For instance, OECT's response could discriminate the low frequency signals if a threshold of 50% of the signal was applied. It was recently proposed that STP effects in neural networks can be modeled as a combination of various frequency filters.[37] Here, the adaptive property of OECTs with in-situ EP represents an attractive feature for neuromorphic synapses engineering that could allow to reproduce complex STP mechanisms. Note that STP effect in our configuration corresponded to

## 3. Conclusion

We report in this paper on the utilization of EP as an in-situ technique for tuning the iono-electronic properties of OECTs. OMIEC optimization has been largely proposed with either volumetric capacitance or effective mobility tuning. We showed here that EP provided a unique technic that allowed controlling both properties simultaneously. This solution is opening new perspectives for the engineering of OECTs where transconductance and impedance can be finely adjusted. More fundamentally, this work is showing how ionic and electronic properties are governing the OECTs responses (see figure S3), which contributes to a better understanding of mix ionic electronic processes in OMIECs.

Along this line, an important bottleneck of OECTs development as bio-sensors is associated with the large variability of soft technologies when circuits and systems need to be designed. Spin-coating as well as inkjet printing, which are the most popular OMIECs fabrication routes, are room-temperature and ambient atmosphere deposition conditions that are prone to high amount of defects in the OMIECs. EP can offer an interesting alternative to adjust the iono-electronic properties of the material post-fabrication to match a tolerable range of variability in these properties. Notably, we showed that standard deviation in transconductance can be reduced by 60%.

This approach is also offering new strategy for the development of adaptive sensors that can adapt to their sensing environment. For instance, electrophysiological recording could benefit from sensors arrays that could be adapted to the specific organization of the biological neural network. Equivalently to learning, adaptive protocols could be designed to adjust OECTs response to match biological neural network organization. It has to be noted that EDOT molecules are non-cytotoxic[40] - at least in the range of concentration reported in this study – and PSS:Na is non acidic. Future work could consider in-situ EP with living cells in the medium. In addition, since electro-active cells / electrode coupling is based on capacitive effect,

OECTs fine-tuning can be used to optimized cells / OECTs coupling or to adjust the filtering of complex bio-signals.

## 4. Experimental Methods

**Materials and instrumentation**. Electropolymerization was done by potentiostatic configuration in an aqueous electrolyte containing 0.1 M of poly(sodium-4-styrene sulfonate) (NaPSS) and 0.01 M of 3,4-ethylenedioxythiophene (EDOT). All chemicals were purchased from Sigma Aldrich. Source and drain as working electrode ($V_{IN}$) and grounded Pt wire as reference electrode ($V_{OUT}$) dipped into the electrolyte.

**Raman**: A 473 nm excitation laser (~1 mW) focused with a 100× objective was used for confocal Raman spectroscopy measurements in air at room temperature. Raman data were treated with Labspec5 software provided by Bruker.

**AFM:** ICON, Bruker atomic force microscopy (AFM), tapping and phase mode images taken in air atmosphere at room temperature ~ 293 K. We used Si cantilevers with a free oscillating frequency $f_0$ ~ 320 kHz and a spring constant k ~ 42 N/m. AFM images were treated with Gwyddion.[38]

**Optical microscopic images** were done by Keyence numerical miscroscope.

**Electrical characterization**. The transistors were characterized using PBS solution as the electrolyte. An Ag/AgCl wire was immersed in the electrolyte and used as the gate electrode. This was the same type of wire that was used as a gate electrode in the Electrochemical impedance spectroscopy (see below). Agilent B1500A semiconductor device analyzer was used to bias the transistor and record the drain current.

**Electrochemical impedance spectroscopy (EIS)** was performed with a Solartron Analytical (Ametek) impedance analyzer from 1 MHz to 1 Hz. All impedance measurements were done in the same electrical ($V_{DC}$ = 100mV and $V_a$ = 20mV) and electrochemical conditions (PBS as electrolyte). Source and drain as working electrode ($V_{IN}$) and grounded Ag/AgCl wire as reference electrode ($V_{OUT}$) dipped into the electrolyte.

**Circuit Impedance Modeling** was performed using an open-source EIS Spectrum Analyzer software.[39] The RC parameter fitting was manually adjusted by simultaneous comparison of the Nyquist plots, Bode's modulus and Bode's phase plots.

.

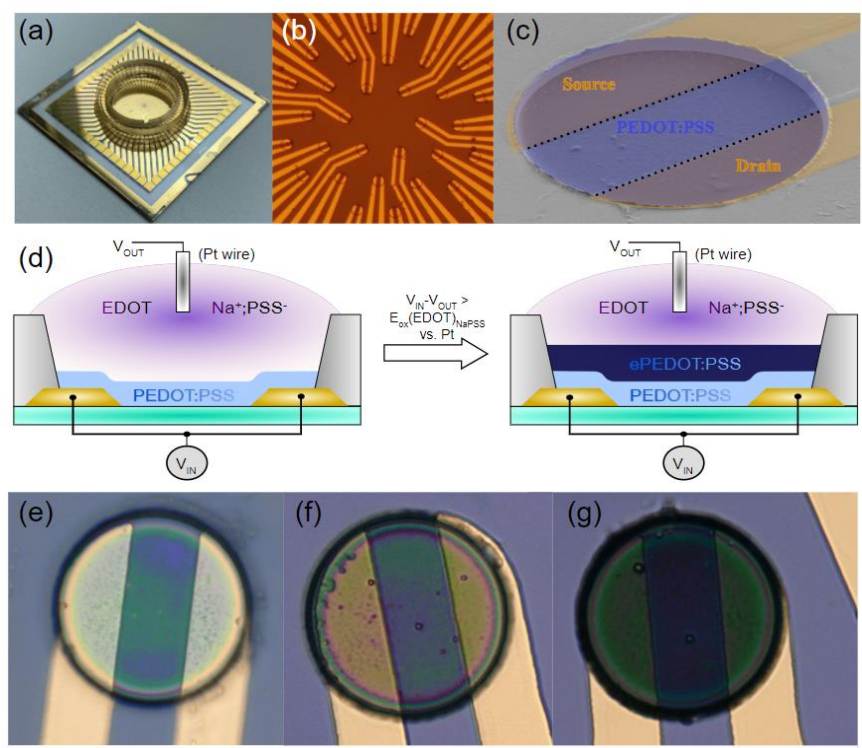

**Figure 1.** Electropolymerization technique for OECTs. (a) OECT glass chip. (b) Microscopic image of the array of 30 OECTs. (c) Colorized SEM image for a PEDOT:PSS OECT. (d) Schematic representation showing the experimental setup to perform potentiostatic electropolymerization of EDOT on top of spin-coated PEDOT:PSS OECT. (e-g) Microscopic images (from left to right) of spin-coated PEDOT:PSS OECT, spin-coated and electropolymerized PEDOT at 0.6 V , spin-coated and electropolymerized PEDOT at 0.7 V.

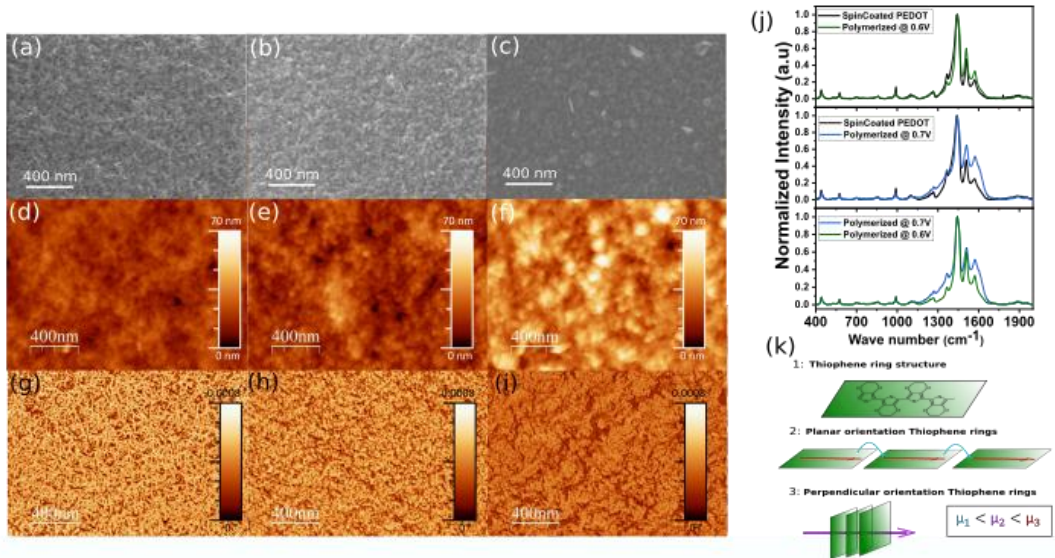

**Figure 2.** Impact of electropolymerization on PEDOT:PSS microstructure.
(a-c) SEM images of the channel's material for spin-coated PEDOT:PSS, electropolymerized PEDOT at 0.6V and electropolymerized PEDOT at 0.7V OECTs. (d-f) AFM height images of the channel's material for spin-coated PEDOT:PSS, electropolymerized PEDOT at 0.6V and electropolymerized PEDOT at 0.7V OECTs. (g-i) AFM phase images of the channel's material for spin-coated PEDOT:PSS, electropolymerized PEDOT at 0.6V and electropolymerized PEDOT at 0.7V OECTs. (j) Typical Raman spectra ($\lambda_{exc}$ = 477 nm) of spin coated PEDOT:PSS (in black), electropolymerized at 0.6 V (in green), and electropolymerized at 0.7 V (in blue). (k) Schematic representation of the PEDOT Thiophene ring structure and its planar and perpendicular orientation structure.

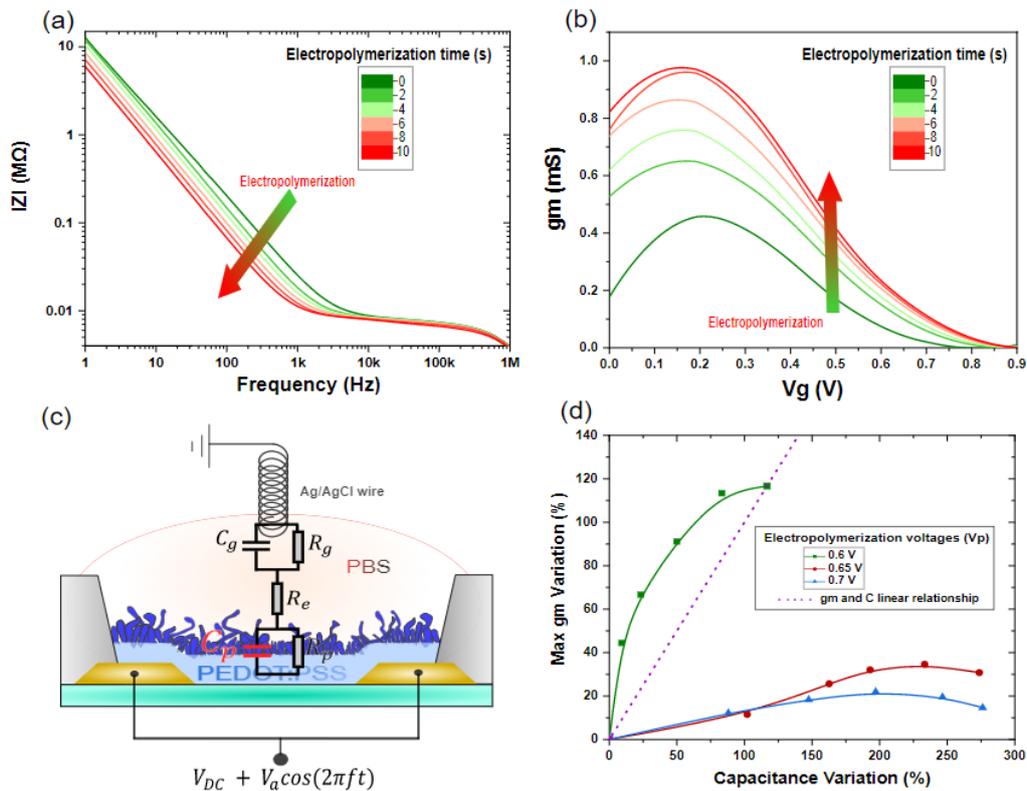

**Figure 3.** Impact of electropolymerization on OECTs' electrical properties.
(a) Bode's diagram of the impedance spectroscopy for an OECT during successive EP steps of 2 s at $V_p = 0.6$ V on top of spin-coated PEDOT:PSS. (b) Effect on the transconductance for the same device as in (a). (c) Equivalent electrical circuit used to extract the total capacitance $C_p$ of the OECT. (d) Correlation graph between the equivalent capacitance (from the impedance spectrogram's fitting) and the device transconductance (from the OECT transfer characteristics) for different EP voltages. Each point from a line are obtained by repeating the same EP conditions for 2 s.

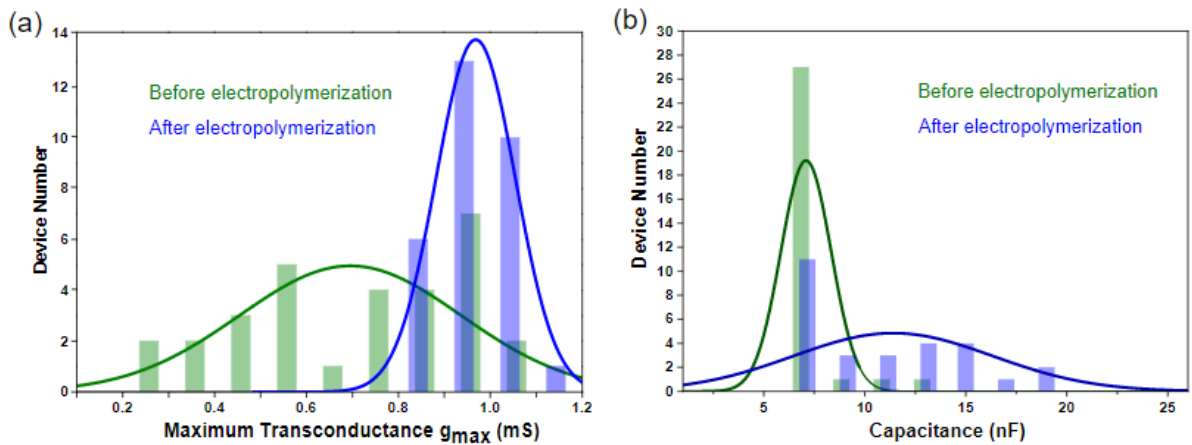

**Figure 4.** Adaptive OECT implementation with EP. (a) green: distribution of the maximum transconductance (mS) for an array of 30 OECTs . An adaptation procedure with target of 1 mS is used to gradually change the transconductance of each OECT sequentially. (blue) Final transconductance reach after device adaptation. (b) Distribution of the capacitance before (green) and after (blue) OECT adaptation. Lines are the Gaussian fitting of the histograms.

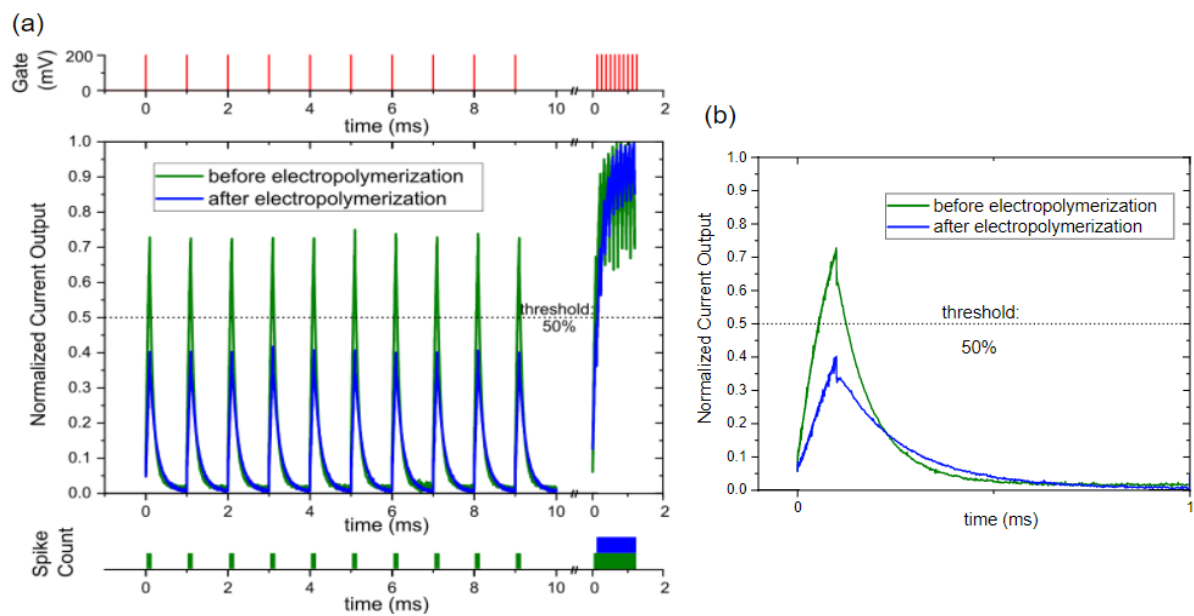

**Figure 5.** Capacitance tuning for adaptive OECTs sensors. (a) The normalized current output response to a train of 100 µs and 200 mV pulses on the gate electrode at 1 KHz and 8 KHz of an OECT before (green) and after (blue) capacitive tuning by EP. Spike-count graph counts the pulses that cross the defined threshold line (50% of output modulation). (b) Normalized current response of an OECT during a single 100 µs pulse.

**Supporting Information**
Supporting Information is available from the Wiley Online Library or from the author.

Acknowledgements

We thank the RENATECH network and the engineers from IEMN for their support.
This work is founded by ERC-CoG IONOS project #773228.

Received: ((will be filled in by the editorial staff))
Revised: ((will be filled in by the editorial staff))
Published online: ((will be filled in by the editorial staff))

References

[1] D. Khodagholy, T. Doublet, P. Quilichini, M. Gurfinkel, P. Leleux, A. Ghestem, E. Ismailova, T. Hervé, S. Sanaur, C. Bernard, G. G. Malliaras, *Nat Commun* **2013**, *4*, 1575.
[2] J. Rivnay, S. Inal, A. Salleo, R. M. Owens, M. Berggren, G. G. Malliaras, *Nat Rev Mater* **2018**, *3*, 17086.
[3] P. Gkoupidenis, N. Schaefer, B. Garlan, G. G. Malliaras, *Adv. Mater.* **2015**, *27*, 7176.
[4] D. A. Bernards, G. G. Malliaras, *Adv. Funct. Mater.* **2007**, *17*, 3538.
[5] C. Yao, Q. Li, J. Guo, F. Yan, I.-M. Hsing, *Adv. Healthcare Mater.* **2015**, *4*, 528.
[6] X. Ji, B. D. Paulsen, G. K. K. Chik, R. Wu, Y. Yin, P. K. L. Chan, J. Rivnay, *Nat Commun* **2021**, *12*, 2480.
[7] J. Y. Gerasimov, R. Gabrielsson, R. Forchheimer, E. Stavrinidou, D. T. Simon, M. Berggren, S. Fabiano, *Adv. Sci.* **2019**, *6*, 1801339.
[8] S. Pecqueur, M. Mastropasqua Talamo, D. Guérin, P. Blanchard, J. Roncali, D. Vuillaume, F. Alibart, *Adv. Electron. Mater.* **2018**, *4*, 1800166.
[9] P. Gkoupidenis, N. Schaefer, X. Strakosas, J. A. Fairfield, G. G. Malliaras, *Appl. Phys. Lett.* **2015**, *107*, 263302.
[10] F. Zenke, E. J. Agnes, W. Gerstner, *Nat Commun* **2015**, *6*, 6922.
[11] D. Khodagholy, J. Rivnay, M. Sessolo, M. Gurfinkel, P. Leleux, L. H. Jimison, E. Stavrinidou, T. Herve, S. Sanaur, R. M. Owens, G. G. Malliaras, *Nat Commun* **2013**, *4*, 2133.
[12] P. R. Paudel, V. Kaphle, D. Dahal, R. K. Radha Krishnan, B. Lüssem, *Adv. Funct. Mater.* **2021**, *31*, 2004939.
[13] A. F. Paterson, A. Savva, S. Wustoni, L. Tsetseris, B. D. Paulsen, H. Faber, A. H. Emwas, X. Chen, G. Nikiforidis, T. C. Hidalgo, M. Moser, I. P. Maria, J. Rivnay, I. McCulloch, T. D. Anthopoulos, S. Inal, *Nat Commun* **2020**, *11*, 3004.
[14] M. N. Gueye, A. Carella, J. Faure-Vincent, R. Demadrille, J.-P. Simonato, *Progress in Materials Science* **2020**, *108*, 100616.
[15] C. M. Proctor, J. Rivnay, G. G. Malliaras, *Journal of Polymer Science Part B: Polymer Physics* **2016**, *54*, 1433.
[16] A. V. Volkov, K. Wijeratne, E. Mitraka, U. Ail, D. Zhao, K. Tybrandt, J. W. Andreasen, M. Berggren, X. Crispin, I. V. Zozoulenko, *Advanced Functional Materials* **2017**, *27*, 1700329.
[17] S. Inal, G. G. Malliaras, J. Rivnay, *Nat Commun* **2017**, *8*, 1767.
[18] A. Giovannitti, D.-T. Sbircea, S. Inal, C. B. Nielsen, E. Bandiello, D. A. Hanifi, M. Sessolo, G. G. Malliaras, I. McCulloch, J. Rivnay, *PNAS* **2016**, *113*, 12017.
[19] P. Schmode, A. Savva, R. Kahl, D. Ohayon, F. Meichsner, O. Dolynchuk, T. Thurn-Albrecht, S. Inal, M. Thelakkat, *ACS Appl. Mater. Interfaces* **2020**, *12*, 13029.
[20] J. Rivnay, P. Leleux, M. Ferro, M. Sessolo, A. Williamson, D. A. Koutsouras, D. Khodagholy, M. Ramuz, X. Strakosas, R. M. Owens, C. Benar, J.-M. Badier, C. Bernard, G. G. Malliaras, *Science Advances* **2015**, *1*, e1400251.
[21] Y. H. Kim, C. Sachse, M. L. Machala, C. May, L. Müller-Meskamp, K. Leo, *Advanced Functional Materials* **2011**, *21*, 1076.
[22] Y. Kim, H. Noh, B. D. Paulsen, J. Kim, I.-Y. Jo, H. Ahn, J. Rivnay, M.-H. Yoon, *Advanced Materials* **2021**, *33*, 2007550.
[23] B. Ji, M. Wang, *J. Micromech. Microeng.* **2020**, *30*, 104001.
[24] B. Ji, M. Wang, C. Ge, Z. Xie, Z. Guo, W. Hong, X. Gu, L. Wang, Z. Yi, C. Jiang, B. Yang, X. Wang, X. Li, C. Li, J. Liu, *Biosensors and Bioelectronics* **2019**, *135*, 181.
[25] S. Garreau, J. L. Duvail, G. Louarn, *Synthetic Metals* **2001**, *125*, 325.
[26] S. Garreau, G. Louarn, J. P. Buisson, G. Froyer, S. Lefrant, *Macromolecules* **1999**, *32*, 6807.
[27] S. Sakamoto, M. Okumura, Z. Zhao, Y. Furukawa, *Chemical Physics Letters* **2005**, *412*, 395.

[28] Q. Zhao, R. Jamal, L. Zhang, M. Wang, T. Abdiryim, *Nanoscale Res Lett* **2014**, *9*, 557.
[29] E. Tamburri, S. Orlanducci, F. Toschi, M. L. Terranova, D. Passeri, *Synthetic Metals* **2009**, *159*, 406.
[30] W. W. Chiu, J. Travaš-Sejdić, R. P. Cooney, G. A. Bowmaker, *Synthetic Metals* **2005**, *155*, 80.
[31] P. V. Almeida, C. M. S. Izumi, H. F. D. Santos, A. C. Sant'Ana, *QuÃ\-mica Nova* **2019**, *42*, 1073.
[32] W. W. Chiu, J. Travaš-Sejdić, R. P. Cooney, G. A. Bowmaker, *Journal of Raman Spectroscopy* **2006**, *37*, 1354.
[33] J. L. Duvail, P. Rétho, S. Garreau, G. Louarn, C. Godon, S. Demoustier-Champagne, *Synthetic Metals* **2002**, *131*, 123.
[34] B. R. Moraes, N. S. Campos, C. M. S. Izumi, *Vibrational Spectroscopy* **2018**, *96*, 137.
[35] N. Sakmeche, S. Aeiyach, J.-J. Aaron, M. Jouini, J. C. Lacroix, P.-C. Lacaze, *Langmuir* **1999**, *15*, 2566.
[36] D. Wu, J. Zhang, W. Dong, H. Chen, X. Huang, B. Sun, L. Chen, *Synthetic Metals* **2013**, *176*, 86.
[37] E. M. Izhikevich, N. S. Desai, E. C. Walcott, F. C. Hoppensteadt, *Trends in Neurosciences* **2003**, *26*, 161.
[38] D. Nečas, P. Klapetek, *Open Physics* **2011**, *10*, 181.
[39] A. S. Bondarenko, G. A. Ragoisha, *J Solid State Electrochem* **2005**, *9*, 845.

[40] S. M. Richardson-Burns, J. L. Hendricks, D. C. Martin, *Journal of Neural Engineering,* **2007**, 6-13.

We show how in-operando elctropolymerization of organic electrochemical transistors can implement adaptive bio-sensors by tuning iono-electronic properties of the material. This work is supported by detail material and device analysis, which bring insightful information on iono-electronic processes for bio-sensing.

M. Ghazal, Dr. M. Daher Mansour, C. Scholaert, Dr. T. Dargent, Dr. Y. Coffinier, Dr. S. Pecqueur, Dr. F. Alibart

**Bio-inspired adaptive sensing through electropolymerization of organic electrochemical transistors**

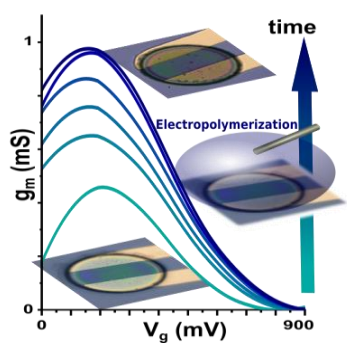

ToC figure

Supporting Information

**Bio-inspired adaptive sensing through electropolymerization of organic electrochemical transistors**

*Mahdi Ghazal, Michel Daher Mansour, Corentin Scholaert, Thomas Dargent, Yannick Coffinier, Sébastien Pecqueur\*, and Fabien Alibart\**

**Table S1**
Assignments of the Raman bands shown in Figure 2

| ν (cm$^{-1}$) | Absorption band |
|---|---|
| 440, 574, 990 | Oxyethylene ring deformation |
| 700 | C-S-C symmetric deformation |
| 1106 | C-O-C deformation |
| 1266 | $C_\alpha$-$C_{\alpha'}$ inter-ring stretching |
| 1366 | $C_\beta$-$C_\beta$ stretching |
| 1444 | $C_\alpha$=$C_\beta$ symmetric vibrations |
| 1508-1570 | $C_\alpha$=$C_\beta$ asymmetric vibrations |

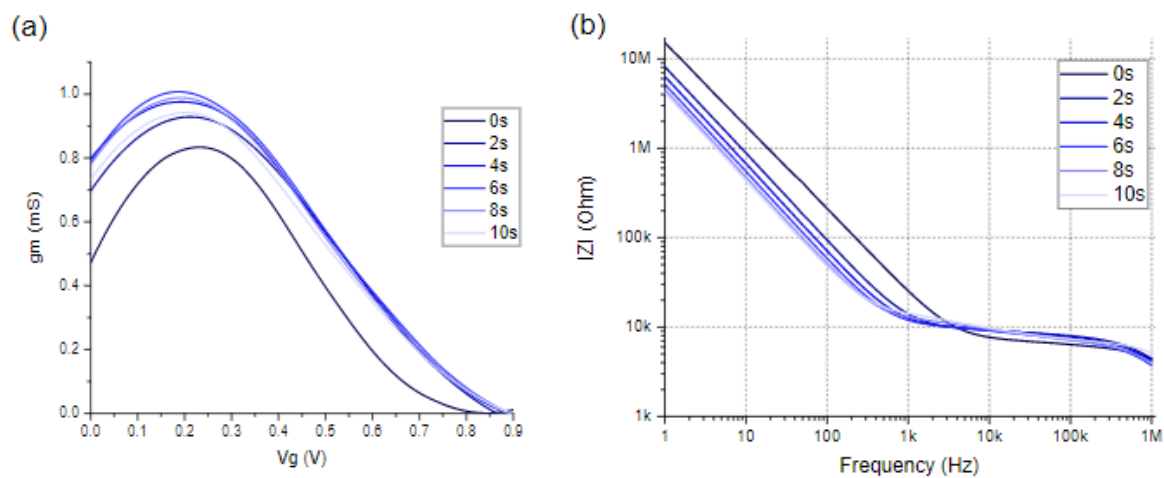

**Figure S1.** (a) Effect on the transconductance of a spin-coated PEDOT:PSS OECT with potentiostatic electropolymerization (2s step EP) at Vp=0.7V. (b) Bode's diagram of the impedance spectroscopy for an OECT upon potentiostatic electropolymerization (2s step EP) on top of spin-coated PEDOT:PSS at Vp=0.7V

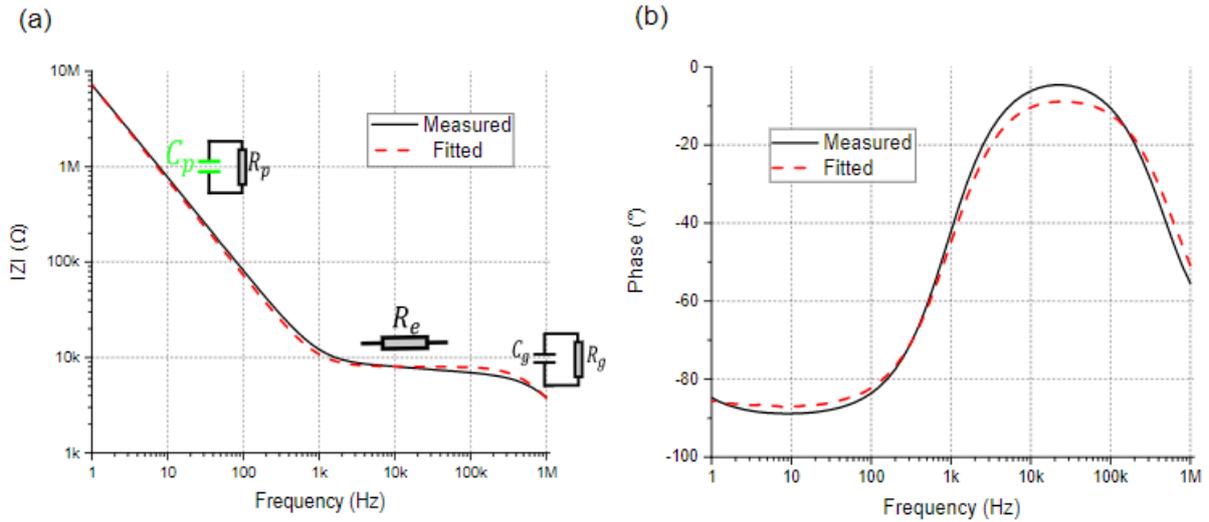

**Figure S2.** (a,b) Bode's and Phase diagram of the impedance spectroscopy of an OECT for measured and fitted examples.

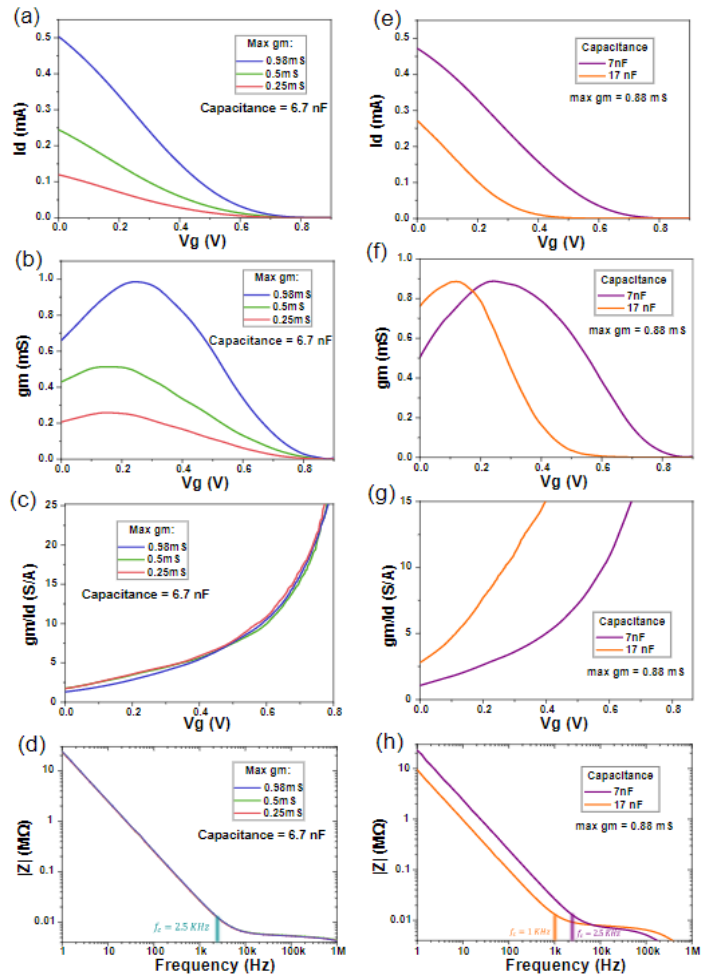

**Figure S3.** How ionic and electronic properties are governing the OECTs responses. (a-d) Transfer characteristics, transconductance (gm), gm/Id and Bode's diagram curves of three different OECTs with same capacitance and different maximum gm values.(e-h) Transfer characteristics, transconductance (gm), gm/Id and Bode's diagram curves of two different OECTs with different capacitance and same maximum gm values.

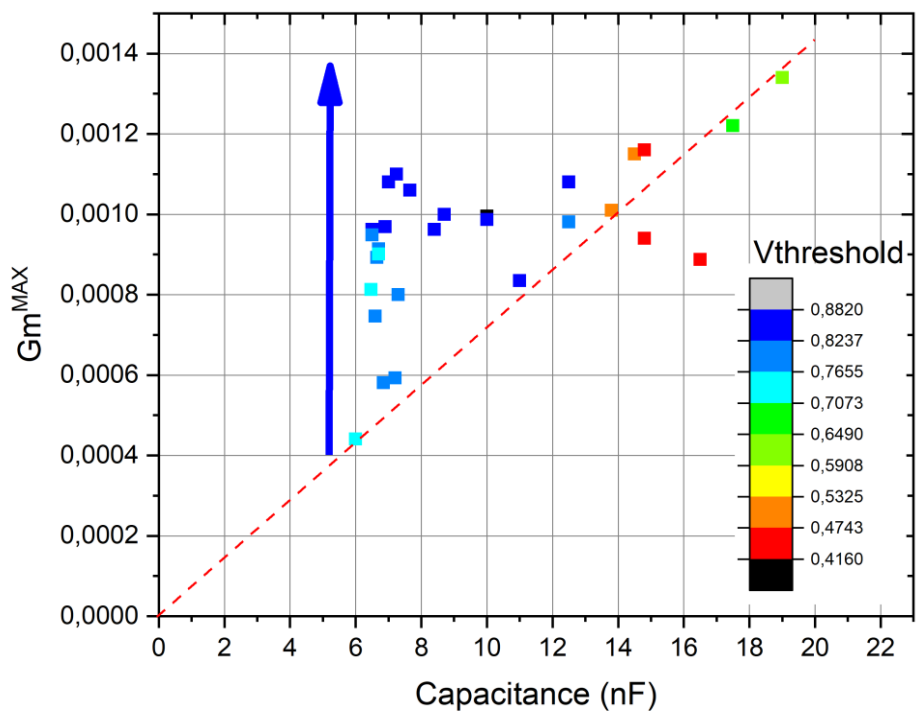

**Figure S4.** Effect of $V_{threshold}$ on tranconductance. The red dashed line indicates the affine relation between tranconductance and capacitance. The blue arrow indicates a change of tranconductance with constant capacitance. The color map of $V_{threshold}$ suggests a larger dependency of $V_{threshold}$ with capacitance whereas an increase of tranconductance with constant C show only a weak dependency with $V_{threshold}$.

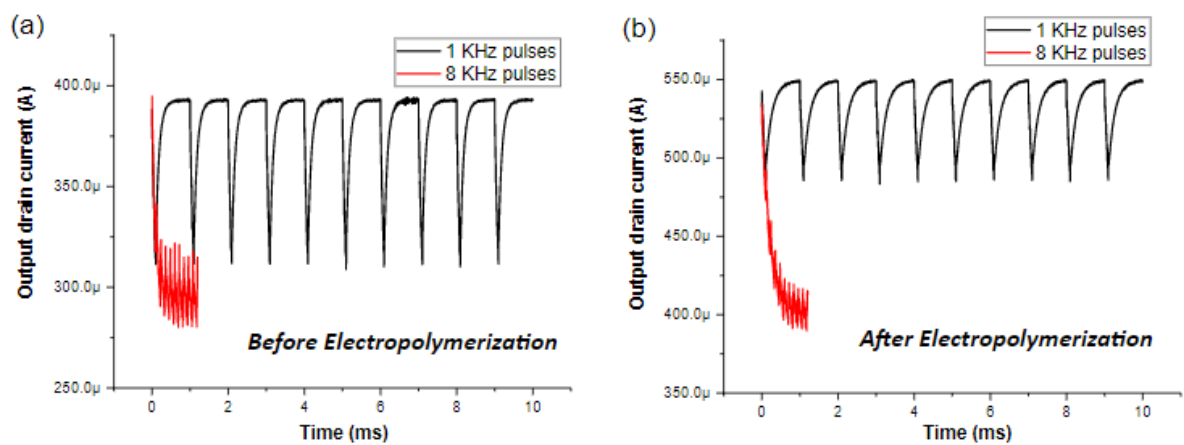

**Figure S5.** Capacitance tuning for adaptive OECTs sensors. The drain current output response (without normalization) to a train of 100 µs and 200 mV pulses on the gate electrode at 1 KHz and 8 KHz of an OECT before (a) and after (b) capacitive tuning by EP.